\documentclass[aps,pre,twocolumn,groupedaddress,showpacs,floatfix]{revtex4}
\usepackage{graphicx}% Include figure files
\usepackage{dcolumn}% Align table columns on decimal point
\usepackage{bm}% bold math
\usepackage{amssymb}
\usepackage{amsmath}
\usepackage{epsfig}
\def\LM#1#2{\left|\begin{array}{l}{#1}\\[1ex]{#2}\end{array}\right.}

\begin{document}
\title{Trapping reactions with subdiffusive traps and particles
characterized by different anomalous diffusion exponents}
\author{S. B. Yuste$^{1}$ and Katja Lindenberg$^{2}$}
\affiliation{$^{(1)}$ Departamento de F\'{\i}sica, Universidad de
Extremadura, E-06071 Badajoz, Spain\\
$^{(2)}$ Department of Chemistry and Biochemistry 0340, and Institute for
Nonlinear Science,
University of California San Diego, 9500 Gilman Drive, La Jolla, CA
92093-0340, USA}
\begin{abstract}
A number of results for
reactions involving subdiffusive species all with the
same anomalous exponent $\gamma$ have recently appeared in the literature
and can often be understood in terms of a subordination principle
whereby time $t$ in ordinary diffusion is replaced by $t^\gamma$.
However, very few results are known for reactions involving different
species characterized by different anomalous diffusion exponents.  Here we
study the reaction dynamics of a (sub)diffusive particle surrounded by
a sea of (sub)diffusive traps in one dimension. We find
rigorous results for the asymptotic survival probability of the particle
in most cases, with the exception of the case of a particle that diffuses
normally while the anomalous diffusion exponent of the traps is smaller
than 2/3.
\end{abstract}

\maketitle
\section{INTRODUCTION}
\label{sec:intro}
In the traditional version of the trapping problem, a normal diffusive
(Brownian) particle ($A$) wanders in a medium doped at random with {\em
static} traps ($B$), and disappears when they meet, $A+B\rightarrow B$.
This problem dates back to Smoluchowski's theory of
reaction rates at the beginning of last century,
and is one of the most widely investigated and applied problems of
non-equilibrium statistical
mechanics~\cite{Hughes,Weiss,ShDba,AvrahamHavlinDifuReacBook}.
An important variation of the basic trapping problem, in which a
diffusive particle wanders in a medium in which the traps are also
diffusive, has been the subject of intense research
since the seminal work of Toussaint and Wilczek~\cite{ToussaintWilczekPRL}.

The principal quantity of interest in the trapping problem is the
survival probability $P(t)$ of the $A$ particles.  From this survival
probability one is able to calculate essentially all other quantities of
practical interest.  Yet this probability is usually difficult to calculate,
and the few instances in which it has been obtained are considered
landmark contributions.  In 1988, Bramson and
Lebowitz~\cite{BramsonLebowitz1,BramsonLebowitz2} proved
rigorously that the long-time survival probability of a
particle diffusing in a one-dimensional medium doped with
diffusive traps decays as $P(t)\sim \exp(-\lambda t^{1/2})$,
$\lambda$ being an undetermined parameter. The evaluation of this
constant proved elusive for many years, engendering much confusion and
proposed solutions that were mutually contradictory. Finally,
quite recently Bray and
Blythe~\cite{BrayBlythePRLPRE} proved in a
simple and elegant way, assuming the so-called ``Pascal principle"
(see below), that the survival probability of a
diffusing particle with diffusion coefficient $D'$ in a
$d$-dimensional medium with $d\le 2$ in which the traps are
also diffusive with diffusion coefficient $D$ is
{\em independent of} $D'$ for long times. They furthermore proved
that the survival probability coincides with
that of an immobile target ($D'=0$) in the
presence of diffusive traps. In particular, in a
one-dimensional medium $P(t)\sim \exp(-4\rho (D
t/\pi)^{1/2})$, where $\rho$ is the density of traps.
Bray and Blythe obtained their results by calculating
an upper and a lower bound for the survival probability that converge to
one another asymptotically.
Some (but not all) of the bounding
results of Bray and Blythe have been extended by Oshanin et
al.~\cite{OshaninEtAlPRE} to systems where the traps
perform a compact exploration of the space, i. e.,
where the fractal dimension $d_w$ of the trajectories of the
traps is greater than the dimension $d$ of the space.

The Pascal principle states that the best strategy for
survival is for the $A$ particle not to move.  This assumption was
adopted in one dimension
``without proof" by Bray and Blythe~\cite{BrayBlythePRLPRE} to calculate
an upper bound
for the survival probability for $d\leq 2$,
although it was already proved in~\cite{Burlatsky}
in the context of incoherent exciton quenching. More recently, the
principle was proved by Bray, Majumdar and
Blythe~\cite{BrayMajumBlythePRE}.
Almost simultaneously but slightly earlier,
Moreau et al.~\cite{MoreauEtAlCondMat} proved
the Pascal principle for a rather general class of random walks on
$d$-dimensional lattices.  These latter authors are responsible for the
appealing name now bestowed on the principle.

The purpose of this paper is to extend the procedure and results of
Bray and Blythe~\cite{BrayBlythePRLPRE}, which are
valid for a Brownian diffusive particle and Brownian diffusive
traps, to situations in which the particle and traps
move subdiffusively.
Anomalous diffusion of a particle is usually characterized
by its mean squared displacement
$x(t)$ for large $t$:
\begin{equation}
\left< x^2(t)\right> \sim \frac{2K_\gamma}{\Gamma(1+\gamma)}
t^\gamma . \label{meansquaredispl}
\end{equation}
Here $K_\gamma$ is the (generalized) diffusion constant and
$\gamma$ is the anomalous diffusion exponent. Ordinary Brownian
diffusion ($\gamma=1$, $K_1\equiv D$) follows Fick's
second law, $\left< x^2(t)\right> \propto t$. The process is
called sudiffusive when $0<\gamma<1$. Subiffusive processes are
ubiquitous in
nature~\cite{MetKlaPhysRep,BouchaudPhysRep90,Kosztolowicz,SubdifuRandPot1,SubdifuRandPot,KantorCM},
and are particularly useful for understanding transport in complex
systems~\cite{ShDba,BouchaudJPI}.

The problem considered in this paper is a special case of
a broad class of reaction-\emph{subdiffusion} processes that have been
studied over the past decades. One approach that has been used to study
these processes is based on the continuous time random
walk (CTRW) with waiting-time distributions between steps that
have broad long-time tails and consequently infinite moments,
$\psi(t)\sim t^{-1-\gamma}$ for $t\to \infty$ with $0<\gamma<1$.
Using the CTRW formalism, Blumen
et al.~\cite{Klafter,BluKlaZuOptical} considered a variety of
reactions including the trapping problem
$A+B(\text{static})\to B(\text{static})$,
the target problem $A\text{(static)}+B\to B$,
and the bimolecular reactions $A+A\to \emptyset$ and $A+B\to \emptyset$.
The moving particles were modeled as continuous-time random walkers with
long-tailed waiting-time densities.  Quite recently, Sung and
Silbey~\cite{SungSilbeyPRL03} have used the CTRW model to study the
dynamics of particles that react at a boundary.  A CTRW approach has
also been applied by
Seki et al.~\cite{SekiJCP1s03} to study the kinetics of
the recombination reaction in subdiffusive media.

Another approach is based on the fractional diffusion equation,
which describes the evolution of the probability density $P(x,t)$
of finding the particle at position $x$ at time $t$ by means of
the fractional partial differential equation (in one
dimension)~\cite{MetKlaPhysRep,SchWysJMP}
\begin{equation}
\frac{\partial }{\partial t} P(x,t)= K_\gamma
~_{0}D_{t}^{1-\gamma } \frac{\partial^2}{\partial x^2} P(x,t)
\label{Pfracdifu}
\end{equation}
where $K_\gamma$ is the generalized diffusion coefficient that
appears in Eq.~(\ref{meansquaredispl}), and
$~_{0}\,D_{t}^{1-\gamma } $ is the Riemann-Liouville
operator,
\begin{equation}
~_{0}D_{t}^{1-\gamma } P(x,t)=\frac{1}{\Gamma(\gamma)}
\frac{\partial}{\partial t} \int_0^t d\tau
\frac{P(x,\tau)}{(t-\tau)^{1-\gamma}}.
\end{equation}
Sung et al.~\cite{SungJCP02} directly addressed
this problem with a fractional diffusion equation approach.
Seki et al.~\cite{SekiJCP1s03} went beyond the CTRW model and
derived a fractional reaction-diffusion equation for the geminate
recombination problem, and one finds some disagreement between the
assumptions and results in this work and that in ~\cite{SungJCP02}.
The fractional diffusion approach
has recently been used to get exact solutions for two types of
one-dimensional
trapping problems: the so called one-sided problem, in which all the
traps lie
on one side of the particle, and the two-sided problem, in which the
traps are located on
both sides of the particle (this is the traditional or standard
version of the trapping problem)~\cite{YusteAcedoSubTrap}.
The fractional-diffusion approach
has also been employed to study other bimolecular reactions between
subdiffusive particles.  In
particular, the annihilation $A+A \to \emptyset$ and coagulation $A+A
\to A$ of subdiffusive particles was
studied~\cite{YusteKatja} by
means of a fractional generalization of the interparticle distribution
function method~\cite{AvrahamHavlinDifuReacBook}. The evolution of
reaction-subdiffusion fronts
for $A+B\to C$ reactions, where both $A$ and $B$ move
subdiffusively, is also
amenable to analysis by means of the fractional diffusion
approach~\cite{YusAceLinSubFront}. Other recent work on fractional
diffusion and CTRW models
of subdiffusive reacting particles can be found in a number of
references~\cite{SubdifuReactOtros}.

In this paper we implement the fractional diffusion equation approach to
study the one-dimensional trapping problem in the
long-time regime for subdiffusive (or diffusive) particles that
move among a distribution of \emph{non-static}
traps. The traps can be either subdiffusive or
diffusive.  For this purpose, we generalize the
ideas of Bray and Blythe~\cite{BrayBlythePRLPRE}.
Recent contributions to the $A+B$ problem based on the fractional diffusion
equation approach~\cite{SungSilbeyPRL03,YusteAcedoSubTrap} share the
simplifying characteristic that the reaction takes place between a
static particle (or fixed boundary) and a subdiffusive particle.
The present work differs from those
in that all the reacting particles (including traps) are
(sub)diffusive and,
moreover, the diffusion constant and the anomalous diffusion exponent of
each species may be different. Some of this work has been presented
in~\cite{ourspie}.

In some cases, asymptotic anomalous diffusion behavior can be found from
corresponding results for normal diffusion with the simple replacement
of $t$ by $t^\gamma$. This can be understood from
a CTRW perpective because
the average number of jumps $n$ made by a subdiffusive walker up to
time $t$
scales as $\langle n \rangle \sim t^\gamma$, and in many instances the
number of jumps is the relevant factor that explains the
behavior of the system.
The simple replacement result is evidence
of ``subordination''
(see Secs. 5 and 7.2 of~\cite{BluKlaZuOptical}).
However, there are other instances where
the behavior of subdiffusive systems cannot be found in this way.
A simple example is the survival probability of subdiffusive
particles in the trapping problem (see Sec. 5
of~\cite{BluKlaZuOptical}).  In particular, for
systems where each species has a different anomalous diffusion
exponent, such a replacement becomes ambiguous.
This is the case for the problem considered here.

Bray and Blythe obtained the asymptotic survival probability of a diffusing
particle in a sea of diffusing traps by calculating an
upper and lower bound that converge asymptotically. We follow their
procedure for subdiffusive particle and traps, with partial success.
While it is possible to obtain convergent bounds in most anomalous
diffusion exponent regimes, this procedure does not work in all regimes.
In particular, the bounding procedure encounters difficulties when the
particle $A$ diffuses normally and the traps are ``too slow" but not
static.

Our paper is organized as follows. In Sec.~\ref{pascal} we extend the
proof of the Pascal principle in one dimension to the case in which both
the particle and the traps move subdiffusively and calculate the upper
bound for the survival probability.
In Sec.~\ref{Sec:upperbound} we present an alternative calculation of
the upper bound, which is particularly helpful for the calculation of
the lower bound  in Sec.~\ref{Sec:lowerbound}.
The survival probability is established, when possible, in
Sec.~\ref{Sec:survival}.  Section~\ref{Sec:panorama}
presents a compendiary of results and some comments on open problems.

\section{THE PASCAL PRINCIPLE}
\label{pascal}
The ``Pascal principle'' of random walks says that the best
survival strategy for a random walker $A$ surrounded by a random sea of
trapping random walkers $B$ is to stand still.
Here we extend the proof of the Pascal principle in one dimension to the
case in which both the
particle and the traps perform subdiffusive random walks.

As did Bray et al.~\cite{BrayMajumBlythePRE}, we consider a finite
volume $V$ containing $N=\rho V$ traps $B$ initially distributed
at random, and a single $A$ particle initially at the origin.
The trajectory of the $A$ particle is $z(t)$.
Bray et al. write the survival probability of $A$ as $P(t)=
\exp\{-\mu[z(t)]\}$, where the trajectory-dependent functional $\mu$ is to
be determined.  To find this functional they derive an equation for it
by calculating, in two ways, the probability density to find
a $B$ particle at the point $z(t)$ at time $t$:
\begin{equation}
\rho =  \int_0^t dt' \dot{\mu}(t') G\left[z(t),t|z(t'),t'\right]
+\rho P\left(\mu(t)\right).
\label{fundamental}
\end{equation}
That the left side is this probability density is obvious.
On the right side $\dot{\mu}(t')dt'= (-\dot{P}/P)dt'$ is the probability
that a $B$ particle intersected $A$ in the time interval $(t',t'+dt')$
for the first time, and the propagator $G$ is
the probability density for this particular $B$ to be at $z(t)$ at
time $t$.
We have slightly augmented Bray et al.'s fundamental starting point
by including the second term on the right hand side,
which is the probability that the first intersection of a $B$ with $A$
occurs at time $t$ and not before.  This term is ultimately unimportant
because it decays much more quickly than the first term and so we omit
it henceforth, but it is satisfying that the fundamental equation now
holds at all times.

Proof of the Pascal principle requires us to show that the trajectory
$z(t)=0$ for all $t$ gives the smallest possible value of $\mu[z]$. For
this trajectory the fundamental equation is
\begin{equation}
\rho = \int_0^t dt' \dot{\mu}_0(t')
G\left[z(t)=0,t|z(t')=0,t'\right],
\label{fundamental2}
\end{equation}
where $\mu_0(t)=\mu[z(t=0)]$.
The propagator $G\left(z(t)=0,t|z(t')=0\right)$ is a function only of $t-t'$,
which we abbreviate as $G(t-t')$, so Eq.~(\ref{fundamental2}) is a
convolution.  Denoting the Laplace transform of $f(t)$ as $\hat{f}(s)$
yields for the transform of (\ref{fundamental2})
\begin{equation}
\hat{\rho}(s)=s\hat{\mu}_0(s)\hat{G}(s),
\label{laplace}
\end{equation}
where we have noted that $\mu_0(t'=0)=0$ because the initial survival
probability is unity.  It follows that
\begin{equation}
\hat{\mu}_0(s) = \frac{\hat{\rho}(s)}{s\hat{G}(s)}.
\label{toinvert}
\end{equation}
Inversion for the subdiffusive case in one dimension gives
\begin{eqnarray}
G(t) &=& \frac{t^{-\gamma/2} }{\sqrt{4\pi K_\gamma}} H_{1,2}^{2,0}\left[0
\LM{(1-\gamma/2 ,\gamma/2)}{(0,1),(1/2,1)} \right]
\nonumber\\
&=&
\frac{t^{-\gamma/2}}{\sqrt{4K_\gamma}\Gamma\left(1-\frac{\gamma}{2}\right)},
\end{eqnarray}
where $\gamma$ is the anomalous diffusion exponent for the traps,
$K_\gamma$ is the associated generalized diffusion constant, and
$H_{1,2}^{2,0}$ is Fox's
$H$-function~\cite{MathaiSaxena,MetzlerKlafter}, whose value at the
given arguments we have used to write the last equality.
The Laplace transform of $G(t)$ is
\begin{equation}
\hat{G}(s)=\frac{s^{\frac{\gamma}{2}-1}}{\sqrt{4K\gamma}}.
\end{equation}
The inverse of Eq.~(\ref{toinvert}) then immediately follows:
\begin{equation}
\mu_0(t) = \frac{\rho \sqrt{4K_\gamma t^\gamma}}{\Gamma
\left(1+\frac{\gamma}{2}\right)}.
\end{equation}

To prove that $z(t)=0$ gives the global minimum of $\mu[z(t)]$ we again
follow Bray et al. and write $\mu=\mu_0+\mu_1$ in
Eq.~(\ref{fundamental}) (without the last term):
\begin{equation}
\rho =  \int_0^t dt'\left[ \dot{\mu}_0(t') +
\dot{\mu}_1(t')\right] G\left[z(t),t|z(t'),t'\right].
\label{fundamental3}
\end{equation}
Adding and subtracting
$(t-t')^{-\gamma/2}/\sqrt{4K_\gamma}\Gamma\left(1-\frac{\gamma}{2}\right)$
to $G$ in the integrand allows cancellation of the left side of
Eq.~(\ref{fundamental3}) against one of the terms on the right, leaving
\begin{eqnarray}
0&=&-\int_0^t dt'
\frac{\dot{\mu}(t')}{\sqrt{4K_\gamma}\Gamma\left(1-\frac{\gamma}{2}\right)}
\frac{R}{(t-t')^{\gamma/2}} \nonumber\\
&&+ \int_0^t dt'
\frac{1}{\sqrt{4K_\gamma}\Gamma\left(1-\frac{\gamma}{2}\right)}
\frac{\dot{\mu}_1(t')}{(t-t')^{\gamma/2}},
\end{eqnarray}
where
\begin{equation}
R=1-\frac{\Gamma\left(1-\frac{\gamma}{2}\right)}{\sqrt{\pi}}
H_{1,2}^{2,0}\left[0
\LM{(1-\gamma/2 ,\gamma/2)}{(0,1),(1/2,1)} \right].
\end{equation}
An explicit expression for $\mu_1$ is obtained by Laplace transforming
this result, solving for $\hat{\mu}(s)$, and inverting,
\begin{align}
\mu_1(t) =&
\frac{1}{\Gamma\left(\frac{\gamma}{2}\right)\Gamma\left(
1-\frac{\gamma}{2}\right)}\int_0^t
\frac{dt_1}{(t-t_1)^{1-\frac{\gamma}{2}}} \notag\\
&\times \int_0^{t_1} dt_2
\frac{R\dot{\mu}(t_2)}{(t_1-t_2)^{\gamma/2}}.
\end{align}

The rest of the argument follows exactly as in Bray et
al.~\cite{ourspie,BrayMajumBlythePRE}. Since $R\geq 0$ and
$\dot{\mu} \geq 0$
(because $P(t)$ is a nonincreasing
function of time), it follows that $\mu_1(t) \geq 0$ for
all paths $z(t)$, with equality when $z(t)=0$ for all $t$.  The survival
probability of our $A$ particle averaged over all possible trajectories
$z$ is $P(t)=\langle e^{-\mu(t)}\rangle_z = \langle
e^{-\mu_0(t)-\mu_1(t)}\rangle_z,$  and the above proof shows that
$\langle e^{-\mu_0(t)-\mu_1(t)}\rangle_z \leq e^{-\mu_0(t)}$.  Therefore,
\begin{equation}
P_U(t)= \exp\left[-\mu_0(t)\right] = \exp\left[-
\frac{\sqrt{4\rho^2K_\gamma t^\gamma}}{\Gamma
\left(1+\frac{\gamma}{2}\right)}\right]
\label{upperresult}
\end{equation}
is a strict upper bound for the survival probability $P(t)$,
thus extending the proof of the Pascal principle to subdiffusive
particles in one dimension.

\section{ALTERNATIVE CALCULATION OF UPPER BOUND FOR THE
SURVIVAL PROBABILITY}
\label{Sec:upperbound}

The survival probability $P_U(t)$
of a static particle $A$ surrounded by a
distribution of mobile traps (the ``target problem")
was considered as early as 1986 using a CTRW
model~\cite{BluKlaZuOptical}, and more recently for
the three-dimensional case by means of a
fractional diffusion approach~\cite{SungJCP02}.
In the previous section we obtained the result~(\ref{upperresult})
following the approach of Bray et al.~\cite{BrayMajumBlythePRE}.
Here we re-calculate $P_U(t)$ by generalizing the original approach of
Bray and Blythe~\cite{BrayBlythePRLPRE} to the subdiffusive
case because it provides results useful for the calculation of the lower
bound in the next section.

Consider a target $A$ of size $2{\mathcal L}$ centered at the origin,
and let ${\mathcal Q}_1(t|y)$ be the probability that the
trap initially placed at
$y>{\mathcal L}$ has not reached the end of the
target at $y={\mathcal L}$ by time $t$. Then, in terms of
the Fox's $H$ function~\cite{MetKlaBoundary},
\begin{align}\label{}
{\mathcal Q}_1(t|y)&=1-H^{10}_{11}\left[\frac{y-{\mathcal L}}
{\sqrt{K_\gamma t^\gamma}}
    \LM{(1 ,\gamma/2)}{(0,1)}   \right] \notag\\
&\equiv
1-H\left[\frac{y-{\mathcal L}}{\sqrt{K_\gamma t^\gamma}} \right].
\end{align}
For $\gamma\rightarrow 1$ the Fox's $H$ function becomes the
complementary error function (with $K_1 \equiv D$),
and the ordinary Brownian motion result is recovered,
\begin{equation}\label{Brownian}
{\mathcal Q}_1(t|y) =
1- \text{erfc}\left(\frac{y-{\mathcal L}}{\sqrt{4 D t}}\right), \qquad
\gamma=1.
\end{equation}
Next consider $N$
independently diffusing traps that are initially placed  at
random in the interval ${\mathcal L}\leq y \leq L+R$. Here
and henceforth $2R$ is the
size of the system, which we will take to infinity at appropriate points
in the calculation. The probability
${\mathcal Q}_N(t)$ that the stationary target $A$ has survived up
to time $t$ is
\begin{align}\label{}
{\mathcal Q}_N(t)&=\prod_{i=1}^N \frac{1}{R}
\int_{\mathcal L}^{{\mathcal L}+R} dy_i
\left\{1-H\left[\frac{y_i-{\mathcal L}}{\sqrt{K_\gamma
t^\gamma}}\right]\right\}\notag \\
&=\left\{1-\frac{1}{R}\int_{\mathcal L}^{{\mathcal L}+R} dy
H\left[\frac{y-{\mathcal L}}{\sqrt{K_\gamma t^\gamma}} \right]\right\}^N,
\end{align}
or, in terms of the density $\rho=N/R$ of traps,
\begin{align}\label{}
{\mathcal Q}_\infty(t)&=\lim_{R\rightarrow \infty}
\left\{1-\frac{1}{R}\int_0^{R} dy
H\left[\frac{y}{\sqrt{K_\gamma t^\gamma}} \right]\right\}^{\rho R}
\notag \\
&= \exp\left\{-\rho \sqrt{K_\gamma t^\gamma} \int_0^{\infty}
dz\, H\left[z \right]\right\}.
\end{align}
Note that the result is independent of the size of the target.
We need to evaluate the integral
\begin{align}\label{}
I_\gamma&= \int_0^{\infty} dz\,H^{10}_{11}\left[z
    \LM{(1 ,\gamma/2)}{(0,1)}   \right],
\end{align}
which can be done from the properties of the Fox's $H$
function~\cite{MathaiSaxena}. One finds that
\begin{equation}\label{}
H^{10}_{11}\left[z
    \LM{(1 ,\gamma/2)}{(0,1)}\right]=\frac{d}{dz} \,H^{10}_{11}\left[z
    \LM{(1+\gamma/2 ,\gamma/2)}{(0,1)}\right].
\end{equation}
But
\begin{equation}\label{}
H^{10}_{11}\left[\infty\LM{(1+\gamma/2,\gamma/2)}{(0,1)}\right]=0
\end{equation}
and
\begin{equation}\label{}
H^{10}_{11}\left[0\LM{(1+\gamma/2,\gamma/2)}{(0,1)}
\right]=\frac{1}{\Gamma(1+\gamma/2)},
\end{equation}
so that $I_\gamma=1/\Gamma(1+\gamma/2)$.
Therefore,
\begin{equation}\label{eq8}
{\mathcal Q}_\infty(t)=\exp\left[-\frac{\sqrt{\rho^2 K_\gamma
t^\gamma}}{\Gamma(1+\gamma/2)}  \right].
\end{equation}
This is the survival probability of the target when all the
traps are located to its right.  When the traps are located on
both sides of the target, the survival probability of the target
is the square of Eq.~\eqref{eq8}:
\begin{equation}\label{eq8b}
    P_U(t)={\mathcal Q}_\infty^2(t)=
\exp\left[-\frac{\sqrt{4\rho^2 K_\gamma
t^\gamma}}{\Gamma(1+\gamma/2)}  \right] .
\end{equation}
This result, which is of course identical to Eq.~(\ref{upperresult}),
is the \emph{upper bound} on the survival probability of the moving
particle.

Incidentally, as is well known, the survival probability for the
target problem is related to the distinct number of sites $S(t)$
visited by a trap up to time $t$~\cite{BluKlaZuOptical,BenichouPRE00},
\begin{equation}\label{eq8c}
P_U(t)=e^{-\rho \langle S(t)\rangle +\cdots},
\end{equation}
where the dots represent higher moments that decay more rapidly with
time.  Comparing this expression with Eq.~\eqref{eq8b}, one
finds that the asymptotic average value $\langle S(t)\rangle$ of
the territory explored up to time $t$ by a subdiffusive walker with
generalized diffusion coefficient
$K_\gamma$ and anomalous diffusion exponent $\gamma$ is
\begin{equation}\label{}
\langle S(t)\rangle \sim \frac{2\sqrt{ K_\gamma
t^\gamma}}{\Gamma(1+\gamma/2)}.
\end{equation}
This result agrees with that found by Yuste and
Acedo~\cite{YusteAcedoSubTrap} using a different approach.

\section{LOWER BOUND FOR THE SURVIVAL PROBABILITY}
\label{Sec:lowerbound}
Let $P_L(t)$ be the  probability that the mobile particle $A$
remains inside a box of
size ${\mathcal L}$ and that all the traps remain outside this box until
time $t$ (we distinguish between ${\mathcal L}$, the size of the
box, and $L$, which denotes ``lower bound").  When this
happens, the particle $A$ survives. It is clear that $P_L(t)$ is a
lower bound for the survival probability $P(t)$ of interest
because there exist many other trajectories involving the
simultaneous presence of the particle and traps within the
box ${\mathcal L}$ that allow the particle $A$ to survive.
This lower bound was first calculated for diffusive particles and traps
by Redner and Kang~\cite{KangRedner} and further considered
(and in some cases corrected) in~\cite{BramsonLebowitz1,Berez}.

The probability $P_L(t)$ is itself the product of
three probabilities:

1. The probability $Q_1$ that at $t=0$ the box of size
${\mathcal L}$ contains no traps:
\begin{equation}\label{notraps}
Q_1=e^{-\rho {\mathcal L}}.
\end{equation}

2. The probability $Q_2$ that no traps enter the box of size
${\mathcal L}$ up to time
$t$: \begin{equation}\label{}
Q_2=\exp\left[-\frac{2}{\Gamma(1+\gamma/2)} \sqrt{\rho^2 K_\gamma
t^\gamma}\right] =P_U(t).
\end{equation}
Note that it is the derivation of the previous section, which explicitly
shows this probability to be independent of the size of the box, that
allows us to write this result.

3. The probability $Q_3$ that the particle has not left the box of
size ${\mathcal L}$ up to time $t$.  We proceed to evaluate this quantity.

Let $W(x,t)$ be the probability of finding the particle $A$ at
position  $x$ at time $t$ if it was at position $x=0$ at time
$t=0$ and there are absorbing boundaries at $x=\pm{\mathcal L}/2$.
Solving the fractional diffusion equation by means of separation
of variables~\cite{MetKlaPhysRep} one finds
\begin{align}\label{}
W(x,t) =&\frac{2}{{\mathcal L}}\sum_{n=0}^\infty (-1)^n \sin \frac{(2n+1)\pi
(x+{\mathcal L}/2)}{{\mathcal L}} \notag\\
&\times E_{\gamma'}\left(-K'_{\gamma'} (2n+1)^2\pi^2
t^{\gamma'}/{\mathcal L}^2\right),
\end{align}
where $K'_{\gamma'}$
and $\gamma'$ are the generalized diffusion constant and the
anomalous diffusion exponent of the particle $A$.  Therefore,
\begin{align}
Q_3&= \int_{-{\mathcal L}/2}^{{\mathcal L}/2} W(x,t) dx \notag\\
&=\frac{4}{\pi}\sum_{n=0}^\infty
\frac{(-1)^n}{2n+1}E_{\gamma'}\left[-K'_{\gamma'} (2n+1)^2\pi^2
t^{\gamma'}/{\mathcal L}^2\right]. \label{Q3a}
\end{align}

Next we distinguish two cases in the handling of the sum in
Eq.~\eqref{Q3a}: first we deal with a
subdiffusive particle, and subsequently with an ordinary diffusive
particle.
In the subdiffusive case, we note that
for large arguments ($z\gg 1$) the Mittag-Leffler function has the
expansion
\begin{equation}\label{}
E_{\gamma'}(-z)=\sum_{m=1}^\infty \frac{(-1)^{m+1}
}{\Gamma(1-\gamma' m) } \,z^{-m}
\end{equation}
so that
\begin{align}\label{}
Q_3= & \frac{4}{\pi} \sum_{m=1}^\infty
\frac{(-1)^{m+1}{\mathcal L}^{2m}}{\Gamma(1-\gamma' m ) \left[\pi^2
K'_{\gamma'} t^{\gamma'}\right]^m} \sum_{n=0}^\infty
\frac{(-1)^n}{(2n+1)^{m+1}}.
\end{align}
Therefore, for $t \rightarrow \infty$ one finds
\begin{align}\label{}
Q_3=  & \frac{1}{8\Gamma(1-\gamma')} \frac{{\mathcal L}^{2}}{K'_{\gamma'}
t^{\gamma'} } +O\left(\frac{{\mathcal L}^{2}}{K'_{\gamma'} t^{\gamma'}
}\right)^2.
\end{align}
Consequently, a lower bound on the survival probability of the particle
$A$ is
\begin{align}\label{}
P_L(t)=&Q_1Q_2Q_3\notag\\
=& e^{-\rho {\mathcal L}}
\exp\left[-\frac{2}{\Gamma(1+\gamma/2)} \sqrt{\rho^2 K_\gamma
t^\gamma}\right] \notag\\
&\times \frac{1}{8\Gamma(1-\gamma')}
\frac{{\mathcal L}^{2}}{K'_{\gamma'} t^{\gamma'} }
\left[1+O\left(\frac{{\mathcal L}^{2}}{K'_{\gamma'} t^{\gamma'}
}\right)\right].
\end{align}
It can easily be ascertained that this expression is maximal
when ${\mathcal L}={\mathcal L}^*\equiv 2/\rho$ (independent of time),
i.e., $P_L(t)\leq P_{L^*}(t)$ with
\begin{align}\label{eq:bound1}
P_{L^*}(t)=& \frac{e^{-2}}{8\Gamma(1-\gamma')}
\left(\frac{2}{\rho}\right)^2 \frac{1}{K'_{\gamma'} t^{\gamma'}
}\notag\\
&\times \exp\left[-\frac{2}{\Gamma(1+\gamma/2)} \sqrt{\rho^2 K_\gamma
t^\gamma}\right] \notag\\
&\times \left[1+O\left(\frac{1}{\rho^2K'_{\gamma'}
t^{\gamma'} }\right)\right].
\end{align}
This then is our best lower bound for the survival probability $P(t)$ of
a subdiffusive particle.

When the particle $A$ diffuses normally, Eq.~\eqref{Q3a} becomes
\begin{align}\label{}
Q_3= &\frac{4}{\pi}\sum_{n=0}^\infty
\frac{(-1)^n}{2n+1}\exp\left[-D'(2n+1)^2\pi^2 t/{\mathcal L}^2\right]
\end{align}
with $D'\equiv K'_1$.  For long times~\cite{BrayBlythePRLPRE}
\begin{align}\label{}
Q_3\sim &\frac{4}{\pi}\exp\left[-D'\pi^2 t/{\mathcal L}^2\right],
\qquad t\gg 1
\end{align}
so that
\begin{align}\label{}
P_L(t)=&Q_1Q_2Q_3 \notag\\
\sim & \frac{4}{\pi} e^{-\rho {\mathcal L}}
\exp\left[-\frac{2}{\Gamma(1+\gamma/2)}
\sqrt{\rho^2 K_\gamma t^\gamma}\right]
\notag\\
&\times \exp\left[-D'\pi^2 t/{\mathcal L}^2\right]
\end{align}
for $t\gg 1$.  This lower bound can again be maximized by
optimizing the value of ${\mathcal L}$.  The optimal value
is~\cite{BrayBlythePRLPRE}
${\mathcal L}^*=\left(2\pi^2 D' t/\rho\right)^{1/3}$ (time dependent),
so that
\begin{align}\label{eq:dominant}
P_L(t)&\leq P_{L^*}(t) \notag\\
&= \frac{4}{\pi} \exp\left[-\frac{2\sqrt{\rho^2 K_\gamma
t^\gamma}}{\Gamma(1+\gamma/2)} -3(\pi^2 \rho^2 D't/4)^{1/3}\right].
\end{align}
Note that
the dominant term inside the bracket depends on the value of $\gamma$,
the anomalous diffusion exponent for the traps. We
distinguish three cases:

1. Traps with $2/3<\gamma\leq 1$. In this case, for $t\gg 1$,
\begin{equation}\label{}
\frac{2\sqrt{\rho^2 K_\gamma t^\gamma}}{\Gamma(1+\gamma/2)} \gg 3(\pi^2 \rho^2
D't/4)^{1/3}
\end{equation}
so that
\begin{equation}
P_{L^*}(t)= \frac{4}{\pi} \exp\left[-\frac{2\sqrt{\rho^2 K_\gamma
t^\gamma}}{\Gamma(1+\gamma/2)} \right].
\end{equation}

2. Traps with $\gamma=2/3$. Now
\begin{align}\label{}
P_L(t)&\leq P_{L^*}(t)\notag\\
&= \frac{4}{\pi} \exp\left[-\left(\frac{2\sqrt{\rho^2
K_\gamma }}{\Gamma(4/3)} -3(\pi^2 \rho^2 D'/4)^{1/3}\right)
t^{1/3}\right],
\end{align}
that is, the second contribution in the exponent in
Eq.~\eqref{eq:dominant} is of the same order as
the first and must thus be retained.

3. Traps with $0<\gamma<2/3$. Now the second term in the
exponent of Eq.~\eqref{eq:dominant} is dominant:
\begin{equation}
P_{L^*}(t)= \frac{4}{\pi} \exp\left[ -3(\pi^2 \rho^2 D't/4)^{1/3}\right].
\end{equation}

In the next section we examine our upper and lower bound results to
establish the behavior of the survival probability of $A$ whenever
possible.

\section{SURVIVAL PROBABILITY}
\label{Sec:survival}
We now combine our upper and lower bound results.
Recall that the label and exponent $\gamma$
is associated with the traps and $\gamma'$ is associated with the
particle $A$.  The upper bound on the survival probability is in all
cases given in Eq.~\eqref{eq8b}, but the lower bound depends on the anomalous
diffusion exponent of the particle. We distinguish the following cases:

1. {\em Subdiffusive particle} ($0<\gamma'< 1$) {\em and diffusive or
subdiffusive traps} ($0<\gamma \leq 1$).  The lower bound is given in
Eq.~\eqref{eq:bound1}, so that $P_{L^*}(t)\leq P(t)\leq P_U(t)$ leads to
\begin{align}
\frac{2}{\Gamma(1+\gamma/2)} \leq & -\frac{\ln P(t)}{\sqrt{\rho^2 K_\gamma
t^\gamma}} \notag \\
&\leq  \frac{2}{\Gamma(1+\gamma/2)}\notag \\
&+ \frac{2\ln\left[\sqrt{\rho^2
K'_{\gamma'} t^{\gamma'}}\right]+2+\ln\left[2\Gamma(1-\gamma')\right]
}{\sqrt{\rho^2 K_\gamma t^\gamma}}  \notag \\
&+O\left(\frac{(\rho^2 K_\gamma t^\gamma)^{-1/2}}
{\rho^2K'_{\gamma'} t^{\gamma'} }\right).
\end{align}
For $t\rightarrow \infty$,
$\ln\left[\sqrt{\rho^2 K'_{\gamma'} t^{\gamma'}}\right]\ll
\sqrt{\rho^2 K_\gamma t^\gamma}$
and the upper and lower bounds converge asymptotically.
We therefore arrive at the explicit asymptotic survival probability
\begin{equation}\label{Ptgral}
P(t)\sim \exp\left[-\frac{2}{\Gamma(1+\gamma/2)} \sqrt{\rho^2
K_\gamma t^\gamma} \right]
\end{equation}
for $0<\gamma\leq 1$ and $0<\gamma'< 1$. Note that for $\gamma=1$ we
recover the normal diffusive result obtained
earlier~\cite{BrayBlythePRLPRE}.
A noteworthy result here is that the survival
probability depends only on the exponent $\gamma$ that characterizes the
traps and not on $\gamma'$ that characterizes the particle.  This is
interesting vis a vis the subordination issue.

2. {\em Diffusive particle} ($\gamma'=1$) {\em and subdiffusive traps
with}
$2/3<\gamma\leq 1$. The bounds here are
\begin{align}
\frac{2}{\Gamma(1+\gamma/2)} &\leq -\frac{\ln P(t)}{\sqrt{\rho^2 K_\gamma
t^\gamma}} \notag\\
&\leq \frac{2}{\Gamma(1+\gamma/2)} + 3
\left(\frac{\pi}{2\rho}\right)^{2/3} \frac{D'^{1/3}}{
 K_\gamma^{1/2} } t^{1/3-\gamma/2},
\label{boundsga1}
\end{align}
and the asymptotic survival probability is again given by Eq.~\eqref{Ptgral}.

3. {\em Diffusive particle} ($\gamma'=1$) {\em and subdiffusive traps
with}
$\gamma=2/3$ (marginal case).
Now $P_{L^*}(t)\leq P(t)\leq P_U(t)$ leads to the more ambiguous
inequalities
\begin{align}
\frac{2}{\Gamma(4/3)} &\leq -\frac{\ln P(t)}{\sqrt{\rho^2 K_\gamma
t^\gamma}} \notag\\
&\leq \frac{2}{\Gamma(4/3)} + 3 \left(\frac{\pi}{2\rho}\right)^{2/3}
\frac{D'^{1/3}}{
 K_\gamma^{1/2}}.
\end{align}
The bounding procedure is therefore not able to predict the value of the
prefactor $\lambda$ in $P(t)=\exp(-\lambda t^{1/3})$, but the
asymptotic behavior $-\ln P(t) \propto t^{1/3}$ is evident.

4. {\em Diffusive particle} ($\gamma'=1$) {\em and subdiffusive traps
with}
$0<\gamma<2/3$.
The bounds here are also given by Eq.~\eqref{boundsga1}, so that the bounding procedure is not able to determine the asymptotic behavior of $P(t)$ at all for this case. We are not even able to assert
the asymptotic stretched exponential
form $P(t)\sim \exp\left(-\lambda t^\beta \right)$.

\section{PANORAMA AND DISCUSSION}
\label{Sec:panorama}
Bray and Blythe~\cite{BrayBlythePRLPRE} have calculated
the asymptotic survival probability of
a diffusive particle $A$ in a randomly distributed sea of
diffusive traps $B$ in one
dimension, and have determined the precise value of the coefficient
$\lambda$ in the classic result $P(t)\sim \exp \left( -\lambda
t^{1/2}\right)$ first obtained by Bramson and
Lebowitz~\cite{BramsonLebowitz1,BramsonLebowitz2}.  Within some
constraints, we have
generalized this result to the case where one or both of the
species move subdiffusively.  Our particle $A$ is
characterized by the anomalous diffusion exponent $\gamma'$ and
generalized diffusion coefficient $K_{\gamma'}$, and the
traps by $\gamma$ and $K_\gamma$. In the process of this generalization,
we have extended the proof of the Pascal principle, that the best
survival strategy of a particle in a sea of moving traps is to remain
stationary, to the case of particles and/or traps that move
subdiffusively.
These may be the first results in the
literature involving two subdiffusive species with different anomalous
diffusion exponents.

When both species are subdiffusive ($\gamma$ and $\gamma'$
both smaller than unity), the
survival probability is independent of $\gamma'$ and determined
entirely by the subdiffusive properties of the traps, cf.
Eq.~\eqref{Ptgral}.  When the particle
moves diffusively ($\gamma'=1$), on the other hand, we are unable to
unequivocally determine the coefficient $\lambda$ for all cases using
this procedure.  If the traps move sufficiently rapidly
($2/3<\gamma\leq 1$) then the result Eq.~\eqref{Ptgral} is still valid.
Note that this reduces to the Bray and Blythe result when $\gamma=1$.
The case $\gamma=2/3$ is marginal in the sense that we can establish the
behavior $P(t)\sim \exp \left( -\lambda t^{1/3}\right)$, but are not
able to determine the constant $\lambda$.
Note that this particular time dependence of the survival probability is
the same as the classic result for the survival probability of a
diffusive particle in a sea of immobile
traps~\cite{donsker}.
If the traps are too slow (``strongly
subdiffusive"), $0<\gamma<2/3$, we are not able
to determine even the time dependence of the survival probability on the
basis of this approach. However, since we find the same
stretched exponential behavior when $\gamma=2/3$ and when $\gamma=0$,
a conjecture as to the behavior throughout
this slow trap regime might be appropriate. The conjecture is that the
survival probability decays as $P(t)\sim \exp (-\lambda t^{1/3})$ in the
entire regime $0\leq \gamma \leq 2/3$.

We thus find that in so far as one can even think of some sort of
subordination principle (and whether such thinking is appropriate
here is debatable), it is determined by the behavior of the
traps, i.e., by the replacement of $t$ by $t^\gamma$.  Even in the range
of exponents where this is possible, it is only possible for the main
asympotic contribution to $P(t)$ but not for the correction
terms to the leading asymptotic term.

It is interesting to note that the value ${\mathcal L}^*=2/\rho$
that maximizes the
lower bound of the survival probability for a subdiffusive particle
$A$ does not grow with time.  This implies that
finite particle size effects could become relevant with increasing
density $\rho$.  This is completely different from the case of a
Brownian particle $A$, since the growth ${\mathcal L}^*\propto t^{1/3}$ now
suppresses such finite size contributions for any given density.

At this point we inject a digression that is relevant not
only to our analysis but also to the original work of Bray and
Blythe~\cite{BrayBlythePRLPRE}.
They assumed that the particle $A$ is initially surrounded by a random
(Poisson) distribution of mobile traps, an assumption also made in our
explicit analysis, cf. Eq.~\eqref{notraps}.  On the other hand,
if at the start of
the observations ($t=0$) the process has already been taking place for
some time $-\tau$ (i.e., if the process started at some time $\tau$ in
the past), then it is known that the distribution around the
surviving particles at time $t=0$ is not of Poisson form.  Those
particles that initially had close-by traps are more likely to have
been trapped already than those that did not, so that those particles
that have survived are surrounded by a region of
fewer than average traps (sometimes referred to as a ``gap").
Bramson and Lebowitz arrive at the conclusion that the configuration of
$B$ particles is nevertheless dominated by a Poisson random
measure~\cite{BramsonLebowitz2}.  In~\cite{ourspie} we confirm
that for any finite $\tau$ the gap does not affect the asymptotic survival
probability results of Bray and Blythe.  The detailed nature of the
gap is different in the diffusive and subdiffusive cases, and unknown in
the latter. However, we
conjecture that it is no more pronouced in the subdiffusive than in the
diffusive system, and that it does not affect our results either.

Our own results of course leave a number of questions unanswered.  One
obvious question concerns the marginal role of the trap
exponent $\gamma=2/3$ when
the particle is diffusive.  Why is this a marginal exponent?  A
connection between this critical value and the fact that for a Brownian
particle the length that maximizes the lower bound of the survival
probability grows as ${\mathcal L}^*\sim t^{1/3}$ seems plausible, but the
conceptual basis for such a relation is not clear.

The most pressing and intriguing puzzle to resolve is that of
calculating the survival probability when the particle $A$ is diffusive
($\gamma'=1$) and the traps are strongly subdiffusive ($0<\gamma<2/3$).
While we conjecture that the survival probability in this regime decays
as $P(t)\sim \exp (-\lambda t^{1/3})$, the upper and lower bounds in this
case do not have the same asymptotic time dependence so we are not able
to test these conjectures on the basis of the procedures used
in this paper.

\section*{Acknowledgments}

This work was partially supported by the Ministerio de Ciencia y
Tecnolog\'{\i}a (Spain) through Grant No. FIS2004-01399, and by the
National Science Foundation under Grant No. PHY-0354937.

\end{document}